# Long-range and local crystal structures of the $Sr_{1-x}Ca_xRuO_3$ Perovskites

Loi T. Nguyen[1], Milinda Abeykoon[2], Jing Tao[3], Saul Lapidus[4], and R.J. Cava[1]

[1]Department of Chemistry, Princeton University, Princeton, New Jersey 08544, USA

[2]National Synchrotron Light Source II, Brookhaven National Laboratory, Upton, New York 11973, USA

[3]Condensed Matter Physics & Materials Science Division, Brookhaven National Laboratory, Upton, NY 11973, USA

[4]X-ray Science Division, Advanced Photon Source, Argonne National Laboratory, Lemont, Illinois 60439, USA

**Abstract**

The crystal structures of the $Sr_{1-x}Ca_xRuO_3$ ($0 \leq x \leq 1$) perovskites are investigated using both long range and local structural probes. High resolution synchrotron powder X-ray diffraction characterization at ambient temperature shows that the materials are orthorhombic to high precision, and we support previous work showing that $Ca^{2+}$ substitution for $Sr^{2+}$ primarily changes the tilting of rigid corner-sharing $RuO_6$ octahedra at their shared oxygen vertices. X-ray pair distribution function analysis for $SrRuO_3$, $CaRuO_3$ and one intermediate composition show them to be locally monoclinic, and no long range or local phase transitions are observed between 80 and 300 K for materials with intermediate compositions. High-resolution transmission electron microscopy shows that the Sr/Ca distribution is random at the nanoscale. We plot magnetic characteristics such as the ferromagnetic Tc, Curie-Weiss theta, effective moment, and ambient temperature susceptibility vs. the octahedral tilt and unit cell volume.

**Introduction**

The ruthenate perovskites, while of long interest, have attracted increased attention since the single layer material $Sr_2RuO_4$ was found to superconduct at 0.93 K [1]. This family is widely viewed as being an ideal system for investigating correlated materials[2],[3],[4],[5],[6],[7], with behavior that ranges from ferromagnetic $SrRuO_3$ to paramagnetic $CaRuO_3$ to superconducting $Sr_2RuO_4$. Calculations of the magnetic properties of $SrRuO_3$ and $CaRuO_3$ argue that the magnetism is governed by the electronic band structure and correlation between Hubbard and Hund's interactions[8],[9]. There are many studies on ferromagnetic $SrRuO_3$ in bulk and thin film form (see, e.g. [10],[11],[12],[13],[14]).

For the $Sr_{1-x}Ca_xRuO_3$ perovskites, the reported orthorhombic symmetry unit cells are larger in volume than simple cubic perovskite cells due to the tilts of the $RuO_6$ octahedra at their shared vertices. (*a* and *c* are square root of 2 larger than a simple cubic perovskite cell while *b* is double the simple cubic perovskite cell value). Although Ca and Sr are both divalent and thus the electron content of the system is unchanged in $Sr_{1-x}Ca_xRuO_3$, the difference in the resulting tilt angles between the rigid $RuO_6$ octahedra due to the accommodation of different size Sr and Ca ions in the perovskite cavities is widely believed to be the structural feature that distinguishes ferromagnetic $SrRuO_3$ from paramagnetic $CaRuO_3$ [15],[16],[17],[18]. Density Functional Theory calculations argue that the larger orthorhombic distortion for $CaRuO_3$ compared to $SrRuO_3$ suppresses the ferromagnetism[19],[20],[21],[22],[23] and $CaRuO_3$ is claimed to be at a critical point between ferromagnetic and paramagnetic states [24],[25],[26]. Many experimental studies of the $Sr_{1-x}Ca_xRuO_3$ perovskite system have been reported [27],[28],[29],[30],[31],[32],[33].

Here we present a high resolution synchrotron powder diffraction study of the long range crystal structures in the $Sr_{1-x}Ca_xRuO_3$ perovskite solid solution, a study of the local crystal structures by X-ray pair distribution function (PDF) analysis, and high resolution transmission electron microscopy (HRTEM) and scanning transmission electron microscopy (STEM) studies of selected members of the series, accompanied by the magnetic characterization of the materials. The synchrotron diffraction study allows us to determine that the ambient temperature structures are orthorhombic to high precision, and the PDF studies show that the materials are locally monoclinically distorted. No structural phase transitions are observed for selected materials down to 80 K. The electron microscopy studies show that the solid solution is highly chemically homogeneous and also the presence of extended structural defects, which may be responsible for the double magnetic hysteresis loops observed for some compositions. The magnetic properties of the solid solution materials show highly systematic behavior and coupled with the structural characterization allow for a correlation of the fundamental magnetic and structural characteristics of the system with the crystal structures. All of the magnetic properties appear to scale simply with the perovskite tilt angles and the unit cell volume.

**Experiment**

High quality single-phase powder samples of $Sr_{1-x}Ca_xRuO_3$ ($0 \leq x \leq 1$) were synthesized by a solid-state method, using $SrCO_3$, $CaCO_3$ and $RuO_2$ (Alfa Aesar > 99.95%) as starting materials. $SrCO_3$ and $CaCO_3$ were dried in an oven at 120°C for 3 days before use. Stoichiometric ratios of the starting materials were mixed thoroughly in a mortar and pestle, transferred to alumina crucibles and heated in air at 1000°C for 24 hours. The powders were reground and heated in air at 1100°C, 1200°C, 1300°C and 1400°C for 12 hours at each temperature. The phase purities were initially determined through laboratory powder X-ray diffraction using a Bruker D8 Advance Eco with Cu Kα radiation and a LynxEye-XE detector. Single-phase powder samples of the $Sr_{1-x}Ca_xRuO_3$ perovskites were then loaded into Kapton sample capillary tubes and synchrotron powder x-ray diffraction characterization was performed at beamline 11-BM at Argonne National Laboratory. The structural refinements were performed with GSAS [34]. The crystal structure drawings were created by using VESTA [35].

Electron diffraction experiments were performed at Brookhaven National Laboratory using a double-Cs corrected JEOL ARM 200F transmission electron microscope (TEM; JEOL, Tokyo, Japan) equipped with multiple sample holders including a double-tilt holder and a Gatan liquid nitrogen sample cooling holder. TEM samples of $Sr_{1-x}Ca_xRuO_3$ (x = 0, 0.3, 0.5 and 1) were prepared directly from the powder material by grinding. Energy dispersive X-ray spectroscopy (EDX) was performed to examine the elemental distributions in different particles.

The X-ray PDF measurements were carried out at beamline 28-ID-1 at NSLS-II, at Brookhaven National Laboratory. The X-ray PDF data on 11 samples in the $Sr_{1-x}Ca_xRuO_3$ solid solution system were collected at room temperature with a 75 keV beam energy and analyzed by using the program PDFgui [36]. Scattering data were normalized to 0.5 second/frame. Temperature-dependent PDF studies for several samples were obtained on the same experimental apparatus.

The magnetic susceptibilities of $Sr_{1-x}Ca_xRuO_3$ powders were measured in a Quantum Design Dynacool PPMS equipped with a VSM option. The magnetic susceptibilities between 1.8 and 300 K, defined as M/H, where M is the sample magnetization and H is the applied field, were measured at the field of H = 1 kOe. The magnetization of the samples as a function of applied field $\mu_0H$ from -9 to 9 Tesla was measured at a temperature of 2 K.

**Results**

**The Long-range crystal structure**

The high signal to noise ratio and high precision of the synchrotron diffraction measurements allow us to determine that all the perovskite samples were single phase within an estimated 1 part in $10^3$ and that all the materials in the $Sr_{1-x}Ca_xRuO_3$ solid solution are orthorhombic to high precision at ambient temperature. All materials crystallize in the orthorhombic space group *Pnma* (No. 62), consistent with previous work [28],[37]. As examples of the changes in the diffraction patterns observed in the solid solution, **Figure 1** shows the shifts of the (200), (121) and (002) peaks as the lattice parameters change from $SrRuO_3$ to $CaRuO_3$. Rietveld refinements of the $CaRuO_3$, $SrRuO_3$ and $Sr_{0.7}Ca_{0.3}RuO_3$ samples are shown as examples in **Figure 2**. Excellent fits are found to the *Pnma* orthorhombic perovskite structural model of the $GdFeO_3$-type structure type, a perovskite with an $a^+b^-b^-$ type tilt system [38]. The insets show a blow-up of the high angle diffraction region, and in specific show the quality of the orthorhombic structure fit at high angles. The lattice parameters, bond angles and bond lengths are listed in **Table 1** for all studied members of the solid solution series.

As shown in **Figure 3**, which plots the cubic perovskite subcell values for *a*, *b* and *c* (orthorhombic *a* and *c* divided by the square root of 2, orthorhombic *b* divided by 2), while both *b* and *c* lattice parameters decrease monotonically as the Ca content in the $Sr_{1-x}Ca_xRuO_3$ perovskite solid solution increases, the lattice parameter *a* does not vary much in comparison and displays a minimum value near *x* = 0.4. Near this composition the material is dimensionally nearly cubic. This general behavior is consistent with one earlier report [32], although a different study [30] reported a slightly different trend of lattice parameters for samples synthesized at 1000°C. (In our hands, a synthetic temperature 1000°C is too low to obtain pure phases; significantly higher temperatures are required.) Based on the structural refinements of the atomic positions against the synchrotron diffraction data, the Ru-O bond lengths do not change significantly within the scatter (**Figure 4a**), and neither do the O-Ru-O bond angles - the angles that describe the shape of the corner-sharing RuO$_6$ octahedra, as shown in **Figures 4c-d**. The slope of the variation of average (Sr/Ca)-O bond length with composition changes subtly between $x = 0.3$ and $x = 0.4$, shown in **Figure 4b**.

The coordination number of M$^{2+}$ (M=Ca/Sr) decreases from 9 to 8 on going from SrRuO$_3$ to CaRuO$_3$, consistent with the relative sizes of the ions. This is illustrated in **Figure 5**, which also includes the equivalent views for the well-known perovskites SrTiO$_3$ and CaTiO$_3$. (The ideal cubic perovskite cavity has 12 equidistant oxygens in a cuboctahedron geometry for an A atom sitting in the center of the perovskite cavity, as is seen for example for SrTiO$_3$ at 300 K.) We note that at equivalent cavity volumes, **Figure 5** shows that the ruthenate perovskites are more distorted than the titanate perovskites. The increasing off-center position of the A site ions can be represented by plotting the variance in A-O bond lengths ($\Sigma((A-O)_i^2-(A-O)_{avg}^2)/(n-1)$) as a function of $x$ (**Figure 6**); the variance increases significantly when going from SrRuO$_3$ to CaRuO$_3$. A subtle crossover point exists where the A atoms on average move to slightly different positions within the perovskite cavity (**Figure 4b**). This crossover is in contrast to the behavior of the cavity volume, which is a smooth function of $x$ (**Figure 7b**). Finally, although it is widely understood that the "shrinkage" of perovskites in which the size of the A site ion decreases in radius is primarily due to the rigid twisting of the octahedra, this can be shown clearly in the current case through a plot of the fraction of total cavity volume, i.e. (total cavity volume of the cell)/(total cell volume), as a function of $x$, (**Figure 7c**) which is very nearly a constant between (84.5 and 85%), though increasing slightly with increasing Ca content ($x$). This is a good indication that the RuO$_6$ octahedra are essentially rigid in this case, while also reflecting that the A site cavity in perovskites occupy a relatively large part of the total cell volume.

When preparing to connect the observed magnetic properties to the observed structural properties, we must first determine how the Ru-O network deviates from the ideal cubic perovskite crystal structure with composition in the solid solution. **Figure 8** thus shows the composition dependence of the perovskite tilt angles - the number of degrees that the Ru-O-Ru angle deviates from the ideal value of 180 degrees where the octahedra share corners. Plots for the high symmetry [001], [110] and [111] directions [39] in the Sr$_{1-x}$Ca$_x$RuO$_3$ solid solution are shown. ([001] tilt $= \phi = \tan^{-1}\left\{\left[4(u_{x(2)}^2 + v_{x(2)}^2)\right]^{\frac{1}{2}}/(a^2 + b^2)^{\frac{1}{2}}\right\}$, [110] tilt $= \theta = \tan^{-1}\left\{\left[4(u_{x(1)}^2 + v_{x(1)}^2)\right]^{\frac{1}{2}}/c\right\}$, and [111] tilt $= \psi = \cos^{-1}(\cos\theta \times \cos\phi)$, where u$_{x(1)}$, u$_{x(2)}$, v$_{x(1)}$ and v$_{x(2)}$ are atomic displacements; a, b and c are lattice parameters). From SrRuO$_3$ to CaRuO$_3$, all three tilt angles increase with composition. The tilt angles plotted have been determined from the internal atomic coordinates [42]. Also in **Figure 9** plotted are a single parameter characterizing the average tilt of the octahedra (taken as the average of the tilt values over the [001] [110] and [111] directions) as a function of both the theoretical (Goldschmidt) and actual observed tolerance factors [39],[40]. (Tolerance factor $\tau = \langle A - O \rangle / (\sqrt{2} \langle B - O \rangle)$, and the theoretical ionic radii were taken from the Shannon table [41]). The variance of the A-O (A=Sr/Ca) bond lengths as a function of the average tilt in Sr$_{1-x}$Ca$_x$RuO$_3$ perovskites is also plotted in **Figure 9b**.

**The Local crystal structure**

The X-ray PDF refinements for the Sr$_{1-x}$Ca$_x$RuO$_3$ perovskites with $x$ = 0, 0.3 and 1 are seen in **Figure 10**. ($G(r) = \frac{2}{\pi} \int_{Q_{min}}^{Q_{max}} Q[S(Q) - 1] \sin QrdQ$, where G(r) is the probability of finding a pair of atoms separated by a distance of r; Q is the magnitude of the scattering vector and Q = 4π sin θ/λ, and S(Q) is the structural function which is the properly corrected and normalized powder diffraction intensity measured from Q$_{min}$ to Q$_{max}$). The local structures are taken as being represented by atomic separations that are less than 5 Å, whereas the average structures are represented by larger separations. For both CaRuO$_3$ and SrRuO$_3$, there are three relatively sharp peaks between 2.3-2.9 Å, corresponding to the presence of three different local M$^{2+}$-O bond lengths (M = Ca or Sr) for the ions in the perovskite cavities. In Sr$_{0.7}$Ca$_{0.3}$RuO$_3$, the smearing of the features in this range implies that the local Sr and Ca positions in the perovskite cavities are impacted by the positions of Ca and Sr in the neighboring cavities [31],[32]. As seen in **Figure 11**, the long range structures determined by synchrotron powder X-ray diffraction refinements of CaRuO$_3$ and SrRuO$_3$ agree well with the long-range structures determined by the X-ray PDF data. The local structures show some deviations, however, shown in **Figure 12**, where lower symmetry monoclinic models fit better than the orthorhombic structural models for distances below 5 Å. These monoclinic structures arise from very small short range distortions (**Table 2**) that are "averaged out" in the long range structures. Finally, the temperature dependent X-ray PDF patterns for Sr$_{0.7}$Ca$_{0.3}$RuO$_3$ and Sr$_{0.6}$Ca$_{0.4}$RuO$_3$, shown in **Figure 13**, reveal no evidence for either local or long-range structural phase transitions between 80 and 300 K. Sharper peaks are observed at 80 K, consistent with expectations for decreasing atomic thermal vibrations with decreasing temperature.

**Figure 14a** shows the ED, EDX, HRTEM and atomic-resolution high-angle annular dark-field (HAADF) STEM (the image intensity is sensitive to the atomic numbers of the elements in this technique) data for $Sr_{0.7}Ca_{0.3}RuO_3$ in the [010] zone (i.e. looking down the crystallographic *b* axis) at ambient temperature, obtained from an area of 150 nm in diameter. The ED data shows bright sharp spots with a uniform spacing of approximately 0.28 nm. EDX, HRTEM and HAADF-STEM confirm that $Sr_{0.7}Ca_{0.3}RuO_3$ is very structurally and chemically homogeneous. In-situ cooling experiments were performed on $Sr_{0.7}Ca_{0.3}RuO_3$ and no structural phase transition was observed in the ED patterns obtained at 90 K (not shown). This is opposite to the case seen in the $La_{0.23}Ca_{0.77}MnO_3$ manganite, which has a dynamic competition between charge-ordered and charge-disordered phases that causes structural phase separation [44]. Similar measurements on a $Sr_{0.5}Ca_{0.5}RuO_3$ sample are also presented in **Figure 14b**. The presence of relatively large structural defects can be seen in the HRTEM image from $Sr_{0.5}Ca_{0.5}RuO_3$, while the areas between the defects remain structurally and chemically homogeneous as shown in the HAADF-STEM image.

**Magnetic properties**

**Figure 15a** shows the magnetic susceptibility for the $Sr_{1-x}Ca_xRuO_3$ perovskites as function of temperature from 1.8 to 300 K - the inverse susceptibility (minus the temperature independent contribution to the susceptibility) is shown in panel **15b** and the raw susceptibility at 300 K is shown in panel **15c**. The latter plot reveals that the susceptibility at high temperatures (i.e. 300 K) decreases systematically as the Ca content in the perovskite increases, i.e. that the material becomes less magnetic with increasing Ca content. The middle panel shows clearly that at high temperature, the magnetic susceptibility follows the Curie-Weiss law with the slopes of the lines, which are due to the value of the magnetic moment, displaying little variation with temperature. The lower panel shows the variation of the total magnetic susceptibility at 300 K with composition. The susceptibility at 300 K is a sum of the temperature independent term, which presumably arise from the presence of non-magnetic electrons in broad electronic bands, and a temperature dependent term, which arises from the electrons that give rise to the magnetism. The temperature-independent term is seen to be essentially independent of composition and the susceptibility is dominated by the magnetic electrons, even at temperatures as high as 300 K for $CaRuO_3$.

The fitting ($\chi = C/(T-\theta_{CW}) + \chi_0$ (where C is the Curie constant, *T* is temperature, $\theta_{CW}$ is the Curie Weiss temperature and $\chi_0$ is the temperature-independent contribution) to the data shown in **b** yields the effective moment ($\mu_{eff} = \sqrt{(8C)}$) and Curie-Weiss temperature as a function of Ca content, plotted in **Figure 16**. The results indicate a linearly decreasing Curie-Weiss temperature when the Ca content in the $Sr_{1-x}Ca_xRuO_3$ solid solution increases, consistent with many previous studies [30], [32],[45],[31],[46],[47],[48]. The sign of the Curie-Weiss theta changes from positive to negative between $x$=0.5 and 0.6, nominally reflecting the shift from dominantly ferromagnetic to dominantly antiferromagnetic interactions in the solid solution. Also, the fits show that the effective magnetic moment per mole increases slightly on going from $SrRuO_3$ to $CaRuO_3$. Comparison to the magnetic data presented in other studies shows that while our data are highly systematic, previous data are not universally so [29], [30].

**Magnetism-structure correlations**

The magnetism-structure correlations found in the current study are summarized in **Figures 17 and 18**. The monotonic variation of magnetic properties with the Ca content in the solid solution is clear, with the figures showing that this variation can be attributed to differences in the octahedral tilts, which change with Ca content and are clearly defined by our structural study. Significantly, the dramatic decrease of the Curie-Weiss theta with Ca content in the $Sr_{1-x}Ca_xRuO_3$ solid solution (**Figure 16b**), and the accompanying decrease in ferromagnetic Tc show a simple variation with both the octahedral tilt and the unit cell volume (**Figure 18**); within error, for both types of correlation, nicely linear behavior is observed. The ferromagnetic Tc of $SrRuO_3$ (when compressed by applied pressure) decreases with unit cell volume via a different trend when compared to the substitution of Ca in $Sr_{1-x}Ca_xRuO_3$ [49]; Tc is significantly higher at any given cell volume. This implies that the magnetic transition from ferromagnetic $SrRuO_3$ to antiferromagnetic $CaRuO_3$ in the $Sr_{1-x}Ca_xRuO_3$ solid solution is not only due to a decrease in the cell volume (and the accompanying octahedral tilt), but is more strongly influenced by other factors, such as the Sr-Ca mixing.

Finally, **Figure 19** shows the magnetization as function of applied magnetic field for $Sr_{1-x}Ca_xRuO_3$ at 2 K. A detail of the data at x = 0.2, 0.3 and 0.4, where double hysteresis loops are observed, is shown in **Figure 19c**. The coercive field and energy products (taken as the integral of the M vs H curve in the first quadrant of the hysteresis loops), shown in **Figure 20**, peak at $x$=0.4, corresponding to the composition where the material is structurally nearly cubic and the lattice parameter $a$ has its lowest value. The remnant magnetization decreases from $SrRuO_3$ to $CaRuO_3$, consistent with the weakening ferromagnetism.

Double magnetic hysteresis loops were clearly observed at $x$=0.3 and 0.4, and hints of the same behavior are seen for some nearby compositions. This unusual behavior was previously observed in this system [32], including in SrRuO$_3$ thin films [12]; the underlying cause has not yet been determined. This phenomenon may or may not be intrinsic, and instead can be a result of the nanoscale structure of the material. Electron microscopy study revealed no nanoscale precipitates in this composition regime, although, in general, extended structural defects were observed that could result in the presence of different kinds of magnetic domain wall pinning and double hysteresis loops. An alternative justification for the double hysteresis loops could be competing interactions between ferromagnetism and anti-ferromagnetism or mixed magneto-crystalline and uniaxial anisotropies in these compounds [50], [51]. How these loops, especially the double ones, are dependent on sample preparation will be of interest for future studies.

**Conclusion**

The average crystal structure of the Sr$_{1-x}$Ca$_x$RuO$_3$ solid solution, studied by synchrotron x-ray diffraction, is determined to be orthorhombic to high precision, supporting earlier studies. The local structure studied by the X-ray PDF method confirms that the average structure is a good description of the materials at longer distances. At shorter distances both SrRuO$_3$ and CaRuO$_3$ appears to be locally monoclinic due primarily to the displacements of the Sr and Ca from the centers of the perovskite cavities. In SrRuO$_3$, Ru-O-Ru angles in the perovskite matrix formed by the corner-sharing RuO$_6$ octahedra are closer to linear than the ones in CaRuO$_3$ but nonetheless deviate significantly from an ideal 180 degree angle present in cubic perovskites. Thus our data support many previous studies showing that both SrRuO$_3$ and CaRuO$_3$ are not cubic perovskites, with the octahedral tilts being significantly larger in the latter material than the former material. The peculiarities of the octahedral tilting in the solid solution make the material look nearly cubic in a small range of Ca contents near $x$=0.4, but this is an artifact because tilts remain present in the octahedral network. The fundamental magnetic characteristics of the system scale linearly with both the octahedral tilt angles and the unit cell volume. We imagine that either case is consistent with electronic structure calculations and previous models for the magnetism. Magnetically, the highest coercive fields and energy products are observed at the composition of the pseudocubic cell near $x$=0.4, but we have not done experiments that allow us to distinguish among the several potential origins for such behavior. High-resolution transmission electron microscopy (HRTEM) and scanning transmission electron microscopy (STEM) characterization show no nanoscale chemical inhomogeneities. Those studies do, however, show the presence of structural defects such as grain or twin boundaries that may be responsible for the observation of double hysteresis loops for some compositions. Further experimental study may involve even more local probes, presumably at the single atom level, of the structure in the solid solution. Also of interest would be high resolution neutron diffraction studies to best quantify the positions of the oxygen atoms within the unit cells. We intentionally do not over-interpret the data obtained, but rather provide it as a basis for future experimental and theoretical study.

**Acknowledgments**

This research was supported by the Gordon and Betty Moore Foundation, grant GBMF-4412. The electron microscopy work was carried out at Brookhaven National Laboratory (BNL) and sponsored by the US Department of Energy (DOE) basic Energy Sciences (BES), by the Materials Sciences and Engineering Division under Contract DE-SC0012704. This research used resources of the National Synchrotron Light Source II, a U.S. Department of Energy (DOE) Office of Science User Facility operated for the DOE Office of Science by Brookhaven National Laboratory under Contract No. DE-SC0012704. Use of the Advanced Photon Source at Argonne National Laboratory was supported by the U. S. Department of Energy, Office of Science, Office of Basic Energy Sciences, under Contract No. DE-AC02-06CH11357. R.C. acknowledges inspiring discussions of this system with theorists Antoine Georges and Andrew Millis.

**Table 1:** Lattice parameters, average bond lengths and bond angles in $Sr_{1-x}Ca_xRuO_3$ system calculated from the GSAS refinement of x-ray synchrotron powder diffraction data.

| x in $Sr_{1-x}Ca_xRuO_3$ | Lattice parameter (Å) | Average Ru-O axial (Å) | Average Ru-O equatorial (Å) | Average Ru-O-Ru axial (°) | Average Ru-O-Ru equatorial (°) | $\chi^2$ |
|---|---|---|---|---|---|---|
| 0 | a = 5.532338(6)<br>b = 7.848885(9)<br>c = 5.571862(5) | 1.988(3) | 1.9842(5) | 162.9 | 162.7 | 1.063 |
| 0.1 | a = 5.53048(2)<br>b = 7.84090(4)<br>c = 5.56060(2) | 1.995(3) | 1.9918(8) | 161.3 | 161.7 | 1.159 |
| 0.2 | a = 5.5261(3)<br>b = 7.8254(4)<br>c = 5.5416(3) | 1.991(3) | 1.9767(1) | 159.9 | 160.4 | 1.962 |
| 0.3 | a = 5.52210(2)<br>b = 7.80917(3)<br>c = 5.52947(2) | 2.015(4) | 1.9766(7) | 159 | 158.7 | 1.750 |
| 0.4 | a = 5.51546(3)<br>b = 7.79265(8)<br>c = 5.50844(5) | 1.983(4) | 1.9827(2) | 157.6 | 157.6 | 1.510 |
| 0.5 | a = 5.51462(3)<br>b = 7.77921(5)<br>c = 5.49551(3) | 1.993(3) | 2.0080(3) | 155.9 | 156.3 | 1.684 |
| 0.6 | a = 5.51913(4)<br>b = 7.78035(8)<br>c = 5.48708(5) | 1.969(3) | 2.0031(2) | 154.8 | 154.5 | 3.478 |
| 0.7 | a = 5.52077(4)<br>b = 7.74664(5)<br>c = 5.45000(4) | 1.994(0) | 1.9779(2) | 154.1 | 153.0 | 1.409 |
| 0.8 | a = 5.52328(3)<br>b = 7.71384(4)<br>c = 5.41858(3) | 1.998(3) | 1.9874(7) | 152.0 | 151.7 | 1.744 |
| 0.9 | a = 5.52767(3)<br>b = 7.68729(4)<br>c = 5.38824(3) | 2.002(3) | 1.9821(8) | 151.7 | 149.5 | 3.420 |
| 1 | a = 5.534338(7)<br>b = 7.662086(9)<br>c = 5.356035(7) | 1.997(2) | 1.9845(6) | 149.7 | 148.7 | 1.876 |

**Table 2**: Atomic parameters for both orthorhombic (*Pnma*) and monoclinic (*P2$_1$/m*) structures of SrRuO$_3$ and CaRuO$_3$ by PDF analysis for atomic separations between 1.5 and 5 angstroms.

SrRuO$_3$-Orthorhombic average structure - space group *Pnma* (No. 62). $A = 5.5405(7)$ Å, $b = 7.8425(4)$ Å, $c = 5.5804(2)$ Å, α= β= γ= 90°.

| Atom | Wyckoff. | Occ. | x | y | z | U$_{iso}$ |
|---|---|---|---|---|---|---|
| Sr | 4c | 1 | 0.01647(2) | ¼ | 0.99522(6) | 0.0057(4) |
| Ru1 | 4b | 1 | 0 | 0 | ½ | 0.0028(6) |
| O1 | 8d | 1 | 0.27292(7) | 0.02941(9) | 0.71061(0) | 0.0139(9) |
| O2 | 4c | 1 | 0.55359(2) | ¼ | 0.07532(9) | 0.0139(9) |

CaRuO$_3$-Orthorhombic average structure - space group *Pnma* (No. 62). $a = 5.5641(0)$ Å, $b = 7.6051(3)$ Å, $c = 5.3689(5)$ Å, α= β= γ= 90°.

| Atom | Wyckoff. | Occ. | x | y | z | U$_{iso}$ |
|---|---|---|---|---|---|---|
| Ca | 4c | 1 | 0.05639(6) | ¼ | 0.98534(2) | 0.0010(4) |
| Ru1 | 4b | 1 | 0 | 0 | ½ | 0.00028(4) |
| O1 | 8d | 1 | 0.29027(2) | 0.05194(7) | 0.69840(3) | 0.0015(2) |
| O2 | 4c | 1 | 0.45932(2) | ¼ | 0.08989(9) | 0.0015(2) |

SrRuO$_3$-Monoclinic local structure - space group *P2$_1$/m* (No. 11), unique axis *b*. $a = 5.5445(6)$ Å, $b = 7.8453(9)$ Å, $c = 5.5825(1)$ Å, α= 90°, β=91.08(8)°, γ= 90°.

| Atom | Wyckoff. | Occ. | x | y | z | U$_{iso}$ |
|---|---|---|---|---|---|---|
| Sr1 | 2e | 1 | 0.01778(1) | ¼ | 0.99479(0) | 0.00349(4) |
| Sr2 | 2e | 1 | 0.51872(0) | ¼ | 0.50476(5) | 0.00360(0) |
| Ru1 | 2b | 1 | ½ | 0 | 0 | 0.00184(7) |
| Ru2 | 2c | 1 | 0 | 0 | ½ | 0.00186(8) |
| O1 | 4f | 1 | 0.27015(9) | 0.02989(2) | 0.72236(1) | 0.01053(4) |
| O2 | 4f | 1 | 0.22179(2) | 0.96011(6) | 0.21718(3) | 0.00936(5) |
| O3 | 2e | 1 | 0.46057(7) | ¼ | 0.05751(6) | 0.02675(5) |
| O4 | 2e | 1 | 0.93558(7) | ¼ | 0.44432(7) | 0.02227(4) |

CaRuO$_3$-Monoclinic local structure - space group *P2$_1$/m* (No. 11), unique axis *b*. $a = 5.5854(9)$ Å, $b = 7.6466(7)$ Å, $c = 5.3284(1)$ Å, α= 90°, β= 91.41(4)°, γ= 90°.

| Atom | Wyckoff. | Occ. | x | y | z | U$_{iso}$ |
|---|---|---|---|---|---|---|
| Ca1 | 2e | 1 | 0.56900(1) | ¼ | 0.50037(6) | 0.00465(1) |
| Ca2 | 2e | 1 | 0.04472(9) | ¼ | 0.97380(0) | 0.00419(8) |
| Ru1 | 2b | 1 | ½ | 0 | 0 | 0.00182(4) |
| Ru2 | 2c | 1 | 0 | 0 | ½ | 0.00188(9) |
| O1 | 4f | 1 | 0.76069(5) | 0.05181(6) | 0.80755(1) | 0.00258(1) |
| O2 | 4f | 1 | 0.30343(1) | 0.45231(9) | 0.68332(9) | 0.00131(0) |
| O3 | 2e | 1 | 0.95290(2) | ¼ | 0.41495(4) | 0.00153(6) |
| O4 | 2e | 1 | 0.45091(1) | ¼ | 0.11188(9) | 0.00227(2) |

**Figure Captions**

**Figure 1**: **Synchrotron diffraction data for materials in the $Sr_{1-x}Ca_xRuO_3$ solid solution**. A selected two-theta range at 300 K is shown to illustrate the effect of the changing unit cell parameters. Peak shifting, followed by overlap and separation are observed in the region of the (200), (121) and (002) reflections as the unit cell parameters change when substituting Ca for Sr.

**Figure 2**: **Representative room temperature Rietveld refinements.** Data and fits for $x = 0$, 0.3, and 1 in $Sr_{1-x}Ca_xRuO_3$. The observed synchrotron X-ray pattern is shown in black, calculated pattern in red, difference in blue, and the pink tick marks denote allowed peak positions. The insets show the details of the data and fits at higher angles.

**Figure 3**: **The perovskite subcells.** The lattice parameters $b$ and $c$ ($b$ divided by 2 and $c$ divided by square root of 2 to normalize them to a standard perovskite subcell) decrease in the perovskite solid solution on going from $SrRuO_3$ to $CaRuO_3$, while the normalized lattice parameter $a$ (normalized in the same fashion as $c$ is normalized) remains relatively constant with a minimum at around $x = 0.4$. Near this composition the material is dimensionally nearly cubic ($a/(\sqrt{2})$ and $c/(\sqrt{2}) \approx b/2$. The standard deviations are smaller than the plotted points in all panels. Inset: The perovskite subcells for $SrRuO_3$ and $CaRuO_3$ ($RuO_6$ octahedra rendered, oxygens shown at the vertices, Sr and Ca balls in the cavities.)

**Figure 4**: **General characterization of the average crystal structures.** (a) The $RuO_6$ coordination polyhedra. (b) The (Sr/Ca)-O bond lengths for the solid solution, showing the crossover in average bond length and change in slope near $x = 0.3$. (c) and (d) the angles and bond lengths within the $RuO_6$ octahedra on going from $SrRuO_3$ to $CaRuO_3$. In $c$ and $d$, the solid lines show the average values. The standard deviations are smaller than the plotted points.

**Figure 5**: **The A-ion coordination environment in $SrTiO_3$, $SrRuO_3$, $CaTiO_3$ and $CaRuO_3$.** There are 12 oxygens surrounding the perovskite cavities, but within that cavity the coordination changes from 9 in $SrRuO_3$ to 8 in $CaRuO_3$. The Ca/Sr-O distances are shown in the red spheres (oxygens). The equivalent views for $SrTiO_3$ and $CaTiO_3$ are also shown. Calculated cage volumes are shown.

**Figure 6: Composition dependence of the A-O bond length variance in the solid solution.** (A = Sr/Ca) for the $Sr_{1-x}Ca_xRuO_3$ perovskites.

**Fig. 7: The composition dependence of the perovskite cell and cavity volumes.** Specification of the (a) cell volume and (b) A-site cavity volume as a function of composition in the $Sr_{1-x}Ca_xRuO_3$ solid solution and (c) their ratio.

**Figure 8**: **The perovskite tilt angles determined from the average structures** (a) The [001], [110] and [111] perovskite tilt angles (taken as the deviation from the ideal 180 degree Ru-O-Ru bond angle) for compounds in the $Sr_{1-x}Ca_xRuO_3$ solid solution. (b) The average perovskite tilt angle as a function of composition. The average tilt is obtained by averaging over the [001] [110] and [111] directions.

**Figure 9: Further characterization of the crystal structures** (a) The average tilt of the $RuO_6$ octahedra in the $Sr_{1-x}Ca_xRuO_3$ perovskites plotted as function of the observed tolerance factor. Inset shows the variation with "theoretical" tolerance factor, and (b) the variance of the A-O bond lengths (A = Sr/Ca) as a function of perovskite tilt.

**Figure 10**: **The X-ray pair distribution function (PDF) data for short distances.** X-ray PDF data of $Sr_{1-x}Ca_xRuO_3$ in the atomic separation range where the short-range structures are dominant. Data for $x = 0$, 0.3 and 1 are shown.

**Figure 11: The X-ray PDF data for $SrRuO_3$ and $CaRuO_3$ for the "average structure" 15-30 angstrom atomic separation range.** Black dots are the data for G(r) and the red lines are the resulting values for G(r) obtained from the synchrotron powder diffraction data structural refinements. The fits are excellent and indicate that the long-range structures of $SrRuO_3$ and $CaRuO_3$ found in the PDF experiments agree well with the average structure determined through the synchrotron diffraction data refinements.

**Figure 12**: **PDF-based local structure comparison of orthorhombic (*Pnma*) and monoclinic ($P2_1/m$) models for $CaRuO_3$ and $SrRuO_3$.** The PDF data for G(r) are the open circles while the fits are shown as red lines. The fits to the locally monoclinic structures ((c) and (d)) are better than the fits to the orthorhombic models obtained from the long range (average) structure refinements. Although $CaRuO_3$ is fit very well, minor discrepancies surprisingly remain for $SrRuO_3$.

**Figure 13**: **Temperature dependent pair distribution functions (G(r)) at lower (*a* and *c*) and higher (*b* and *d*) interatomic distances for $Sr_{0.6}Ca_{0.4}RuO_3$ and $Sr_{0.7}Ca_{0.3}RuO_3$.** Data at 80, 200 and 300 K are displaced along the vertical axis for clarity. There are no structural phase transitions either locally or on the long range between 80 and 300 K.

**Figure 14**: **Transmission Electron Microscope Characterization at the nanoscale.** Left panel: (a) An electron diffraction pattern of $Sr_{0.7}Ca_{0.3}RuO_3$ in the [010] zone, obtained from an area about 150 nm in diameter. (b) A typical EDX spectrum of $Sr_{0.7}Ca_{0.3}RuO_3$. (c) A HRTEM image and (d) a HAADF-STEM image of $Sr_{0.7}Ca_{0.3}RuO_3$ (in the [010] zone). Right panel: (e-h) Similar measurements on $Sr_{0.5}Ca_{0.5}RuO_3$. Defects can be seen in the crystal structure in the (g) HRTEM image. (h) The AHAADF-STEM image shows the typical atomic arrangements in the *a-b* plane in areas without defects, in addition to the extended defects.

**Figure 15**: **The magnetic characterization of the perovskites in the $Sr_{1-x}Ca_xRuO_3$ system.** (a) Magnetic susceptibility, (b) its inverse as function of temperature, and (c) the observed magnetic susceptibility at 300 K; both the total susceptibility and the temperature independent part of the susceptibility are shown in panel c.

**Figure 16**: **Composition dependence of the magnetic properties.** (a) Effective moment and (b) Curie-Weiss temperature as functions of $x$ in the $Sr_{1-x}Ca_xRuO_3$ perovskites. A comparison between our work (solid symbols) and prior studies Ref. 29 (orange squares) and Ref. 30 (blue circles) is shown.

**Figure 17: Magnetic characterization at high temperature as a function of average perovskite tilt angle and cell volume in $Sr_{1-x}Ca_xRuO_3$.** (a) The magnetic susceptibility at high temperature (300 K) and (b) The effective magnetic moment per Ru. (c) and (d) analogous plots of the magnetic characteristics against the unit cell volume.

**Figure 18: The Curie-Weiss temperature and temperature of transition to the ferromagnetic state in $Sr_{1-x}Ca_xRuO_3$** (a) as a function of perovskite tilt angle and (b) as a function of the unit cell volume. Shown for comparison in panel (b) is the data for the change in the ferromagnetic Tc of SrRuO₃ under applied pressure, taken from reference 49.

**Figure 19**: **The magnetic hysteresis loops.** (a-b) Magnetization as function of applied magnetic field in $Sr_{1-x}Ca_xRuO_3$ at 2 K. (c) A detail of the data for $x$ = 0.2, 0.3 and 0.4.

**Figure 20**: **Composition dependence of the energy product (M*H) for $Sr_{1-x}Ca_xRuO_3$.** The data employed are from the first quadrant of the hysteresis loops. The insets show the coercive fields ($H_c$) and remnant magnetizations ($M_R$) for all compositions.

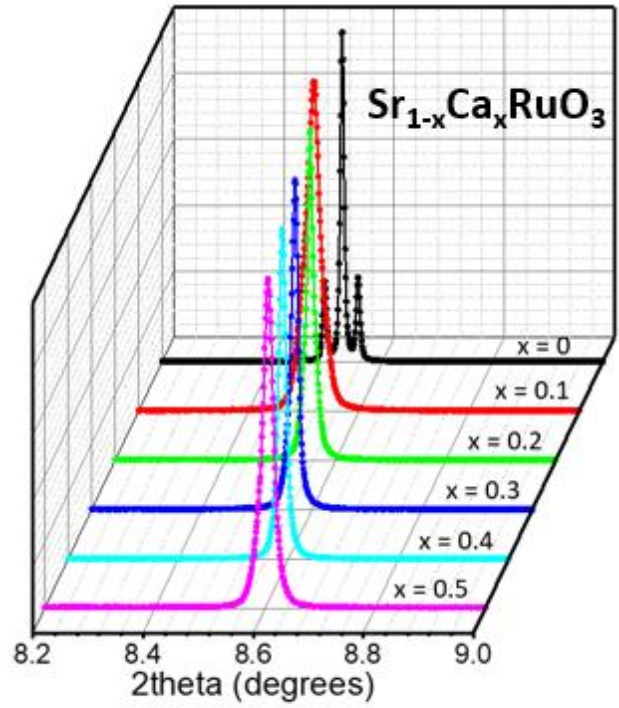
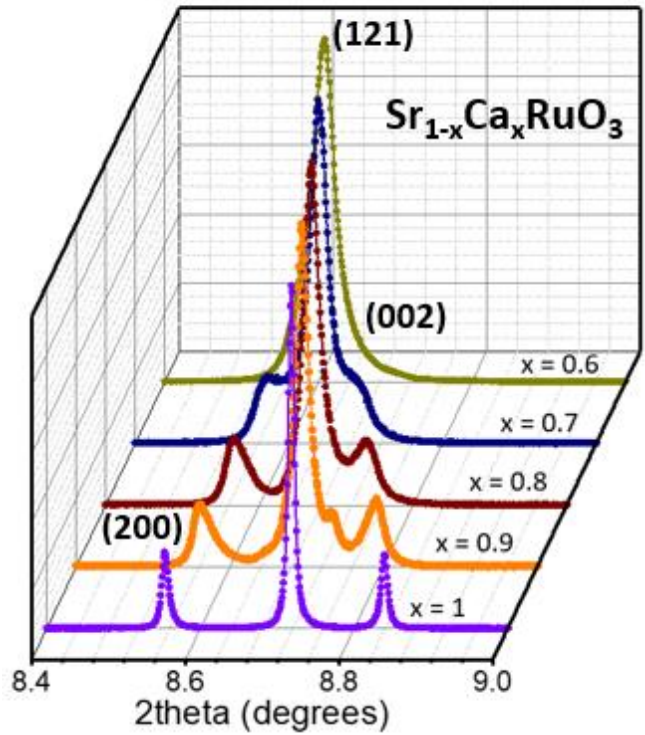

**Fig. 1**

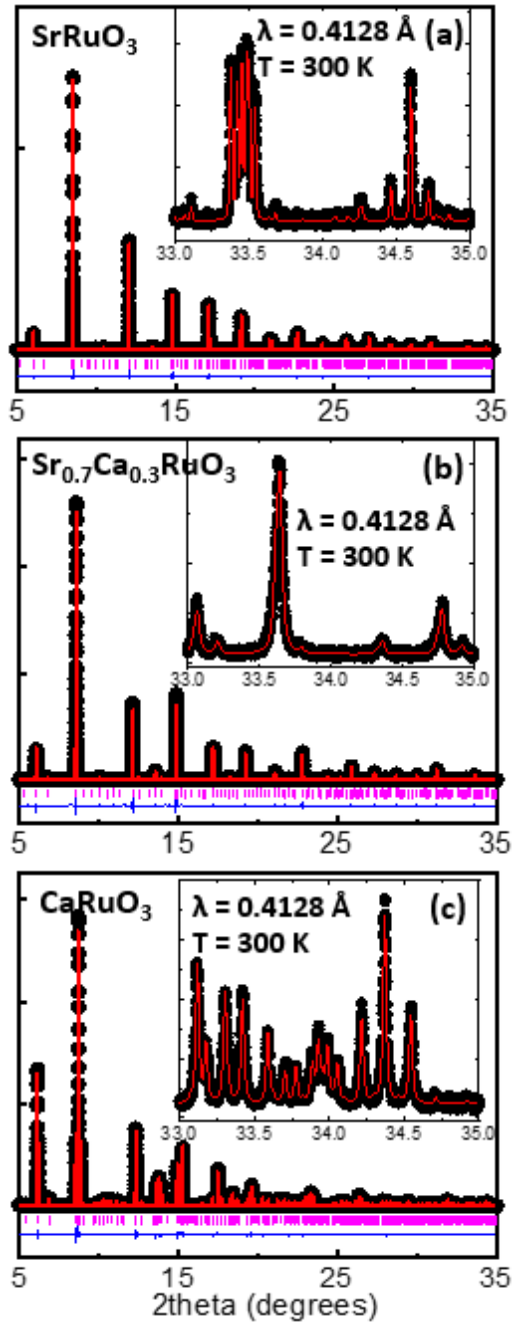

**Fig. 2**

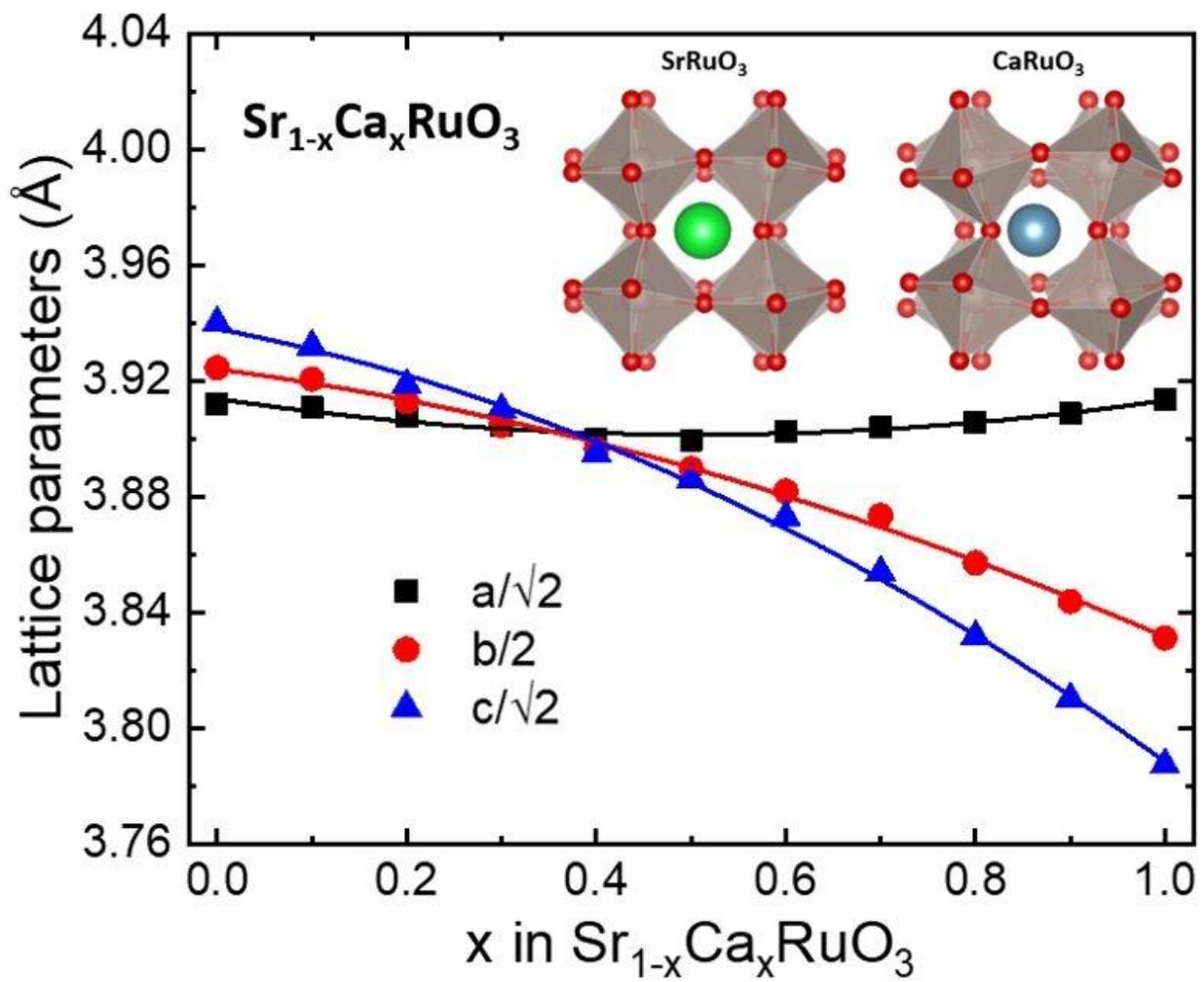

**Fig. 3**

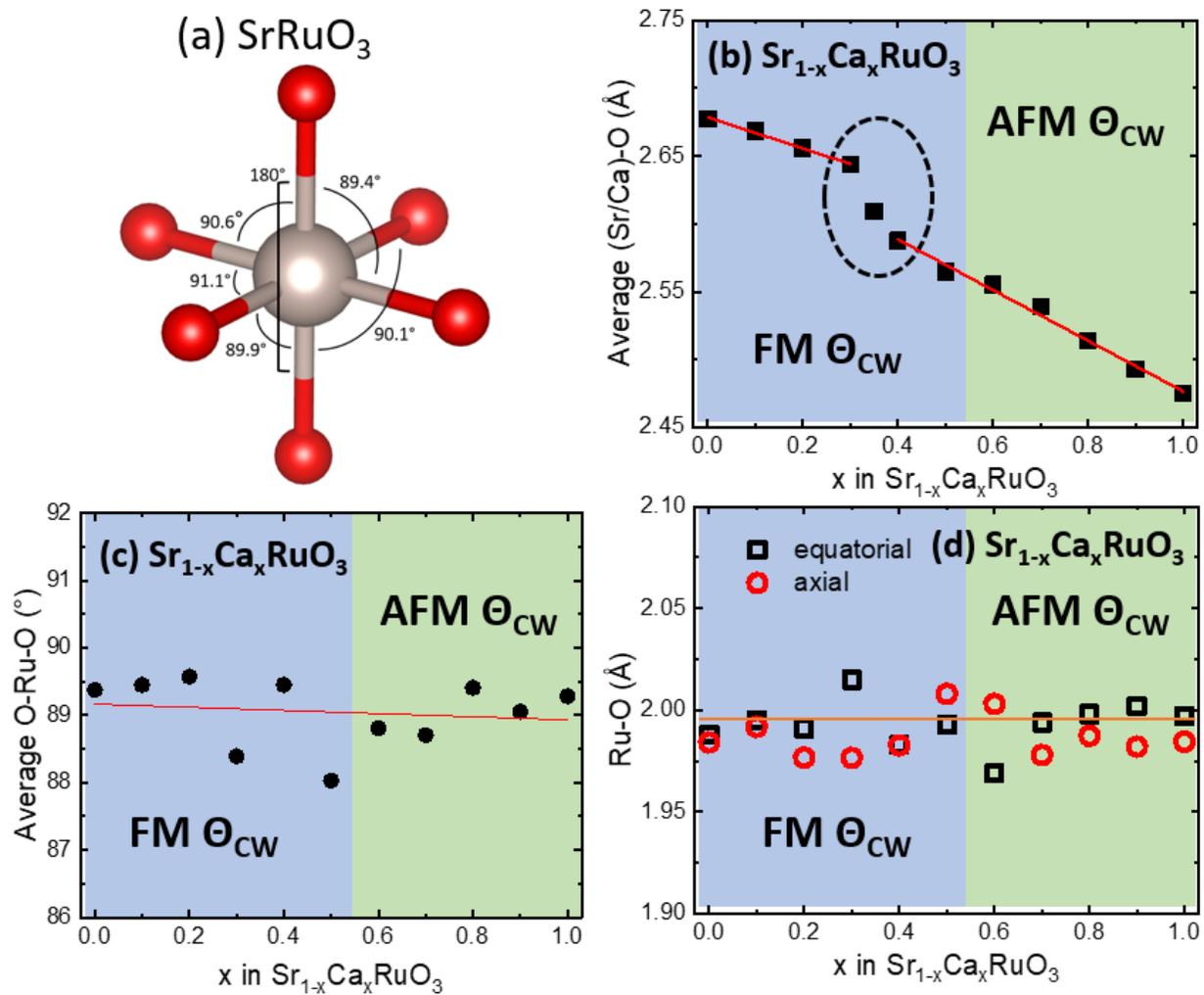

**Fig. 4**

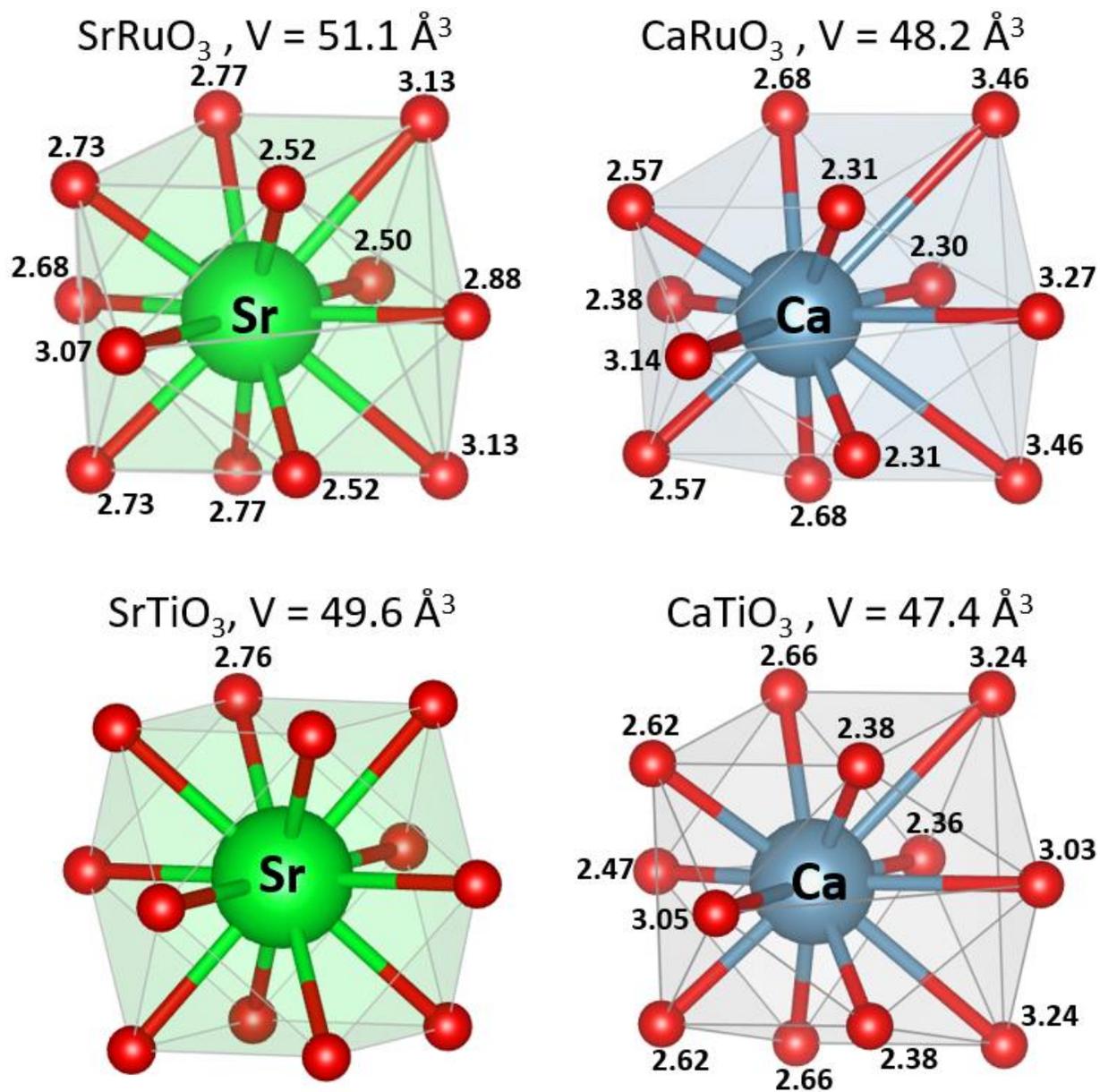

**Fig. 5**

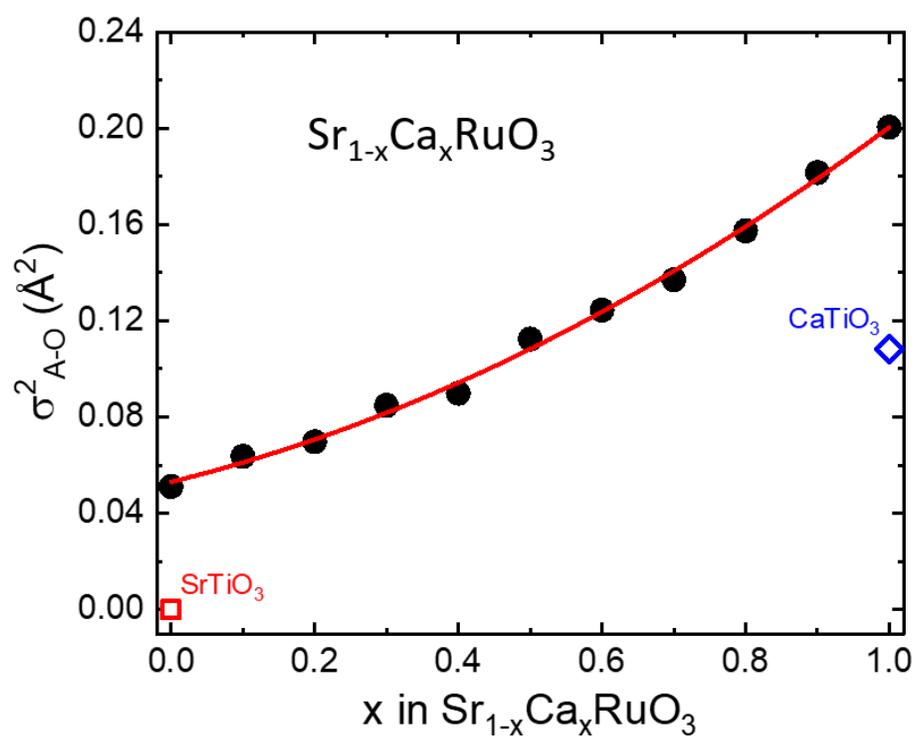

**Fig 6**

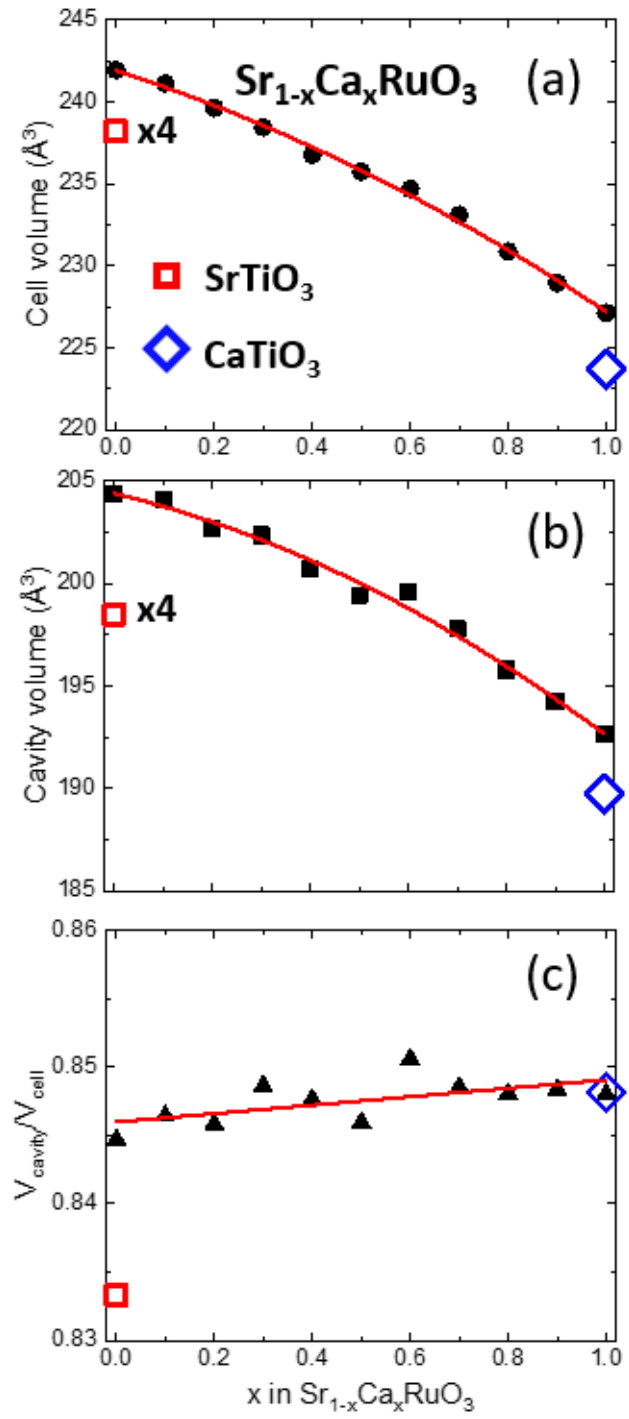

**Fig 7**

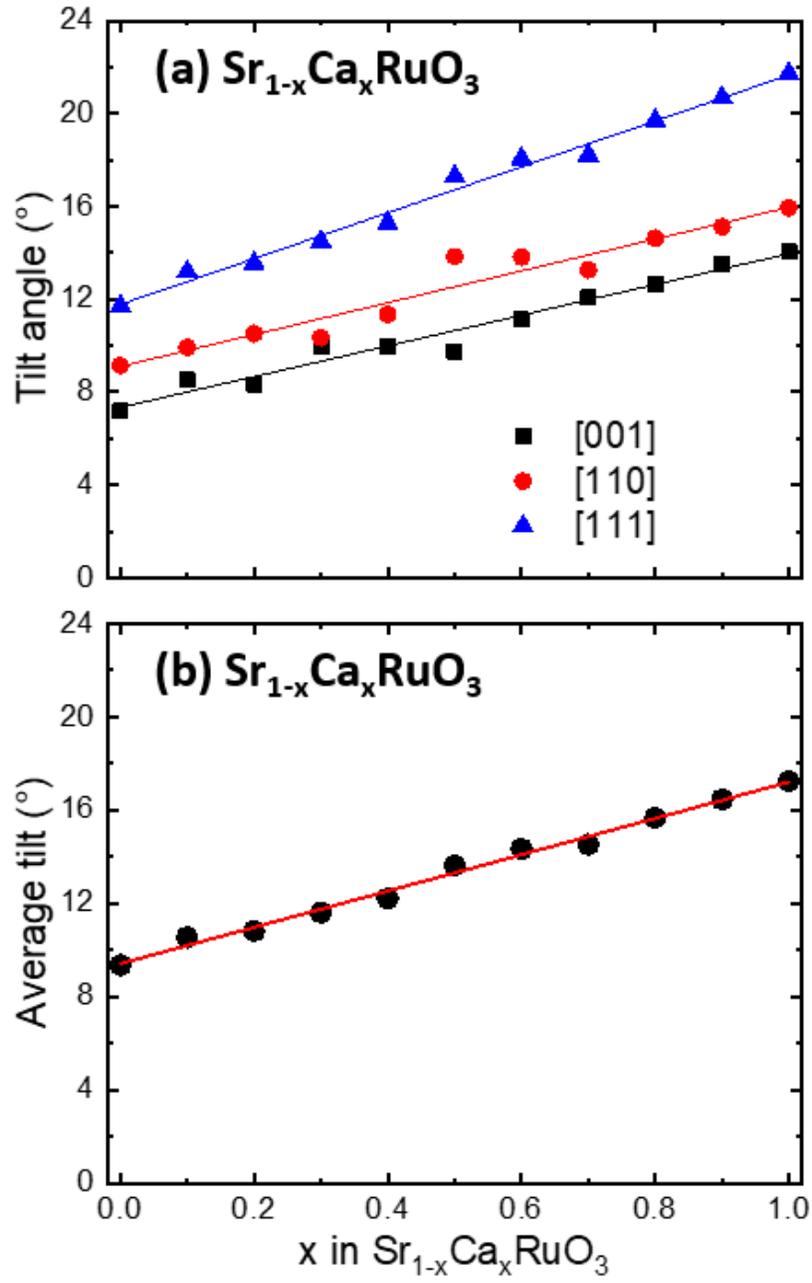

**Fig 8**

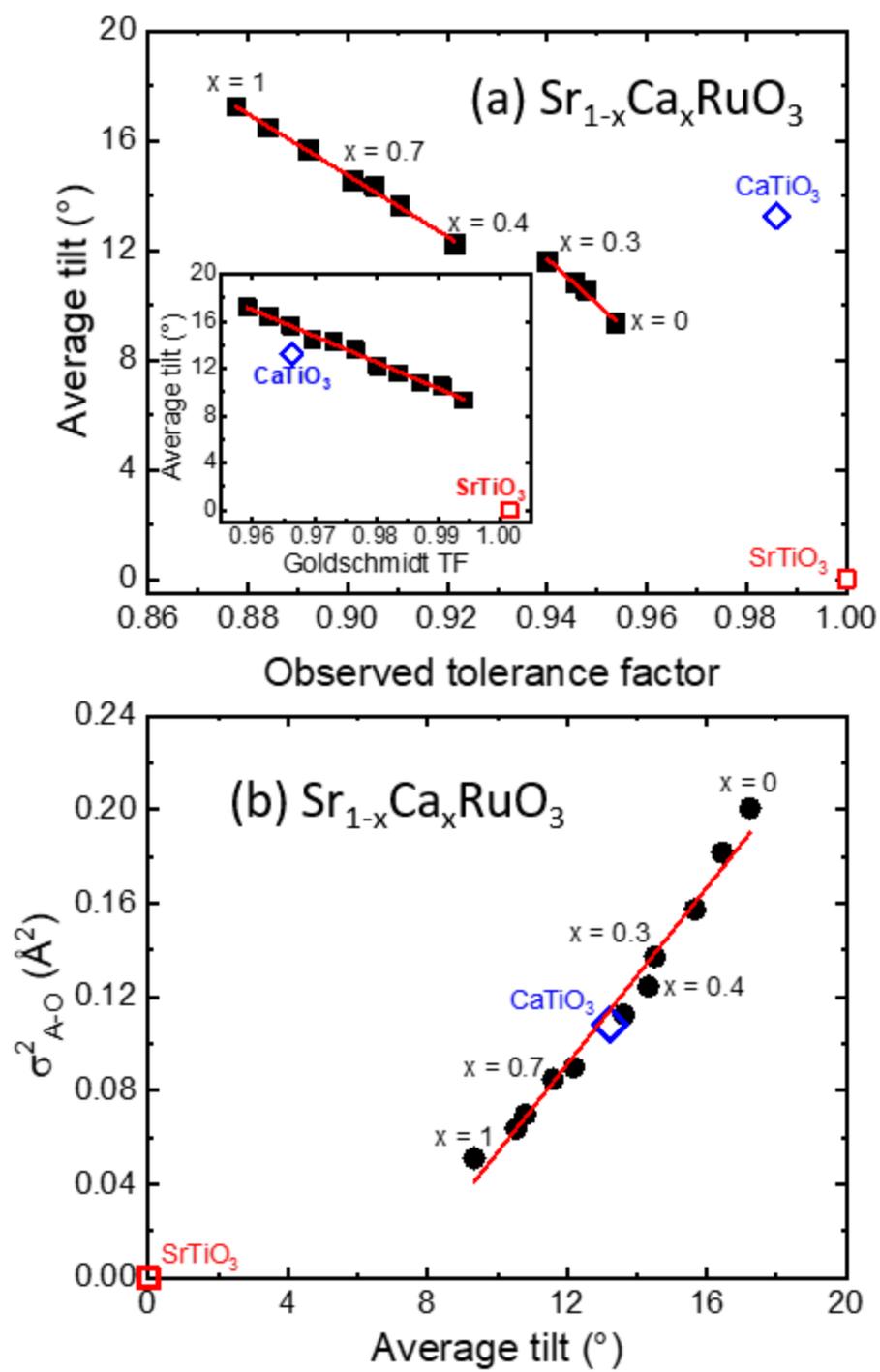

**Fig 9**

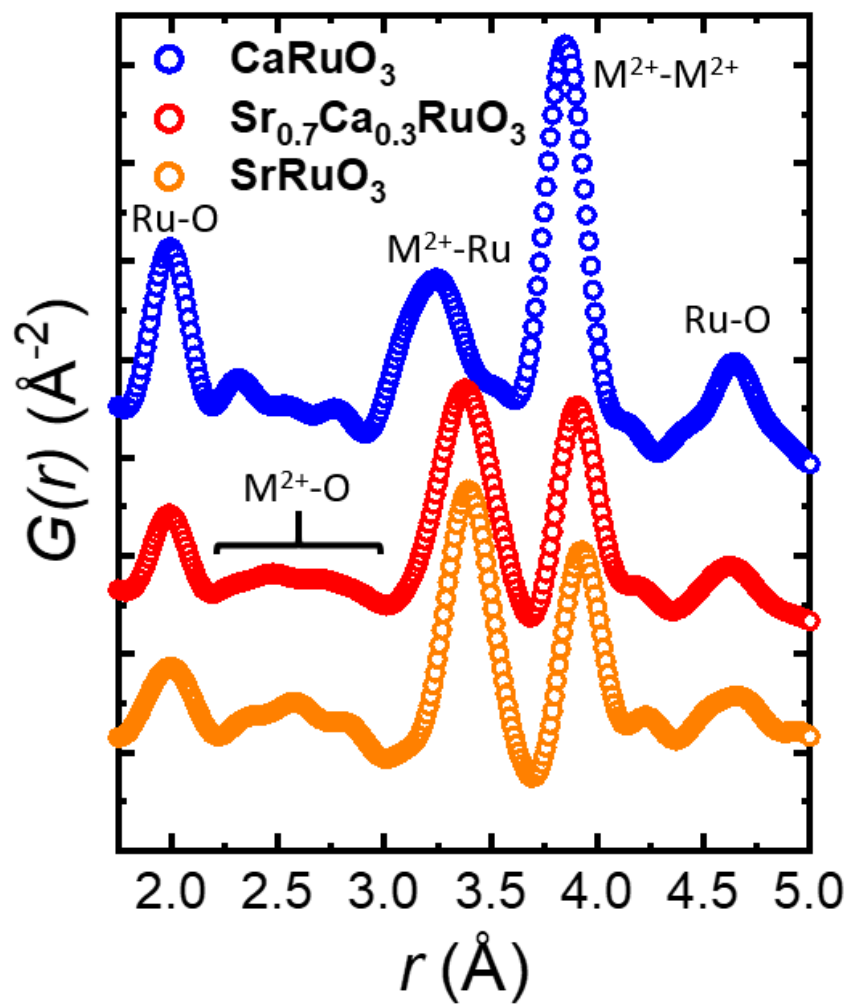

Fig. 10

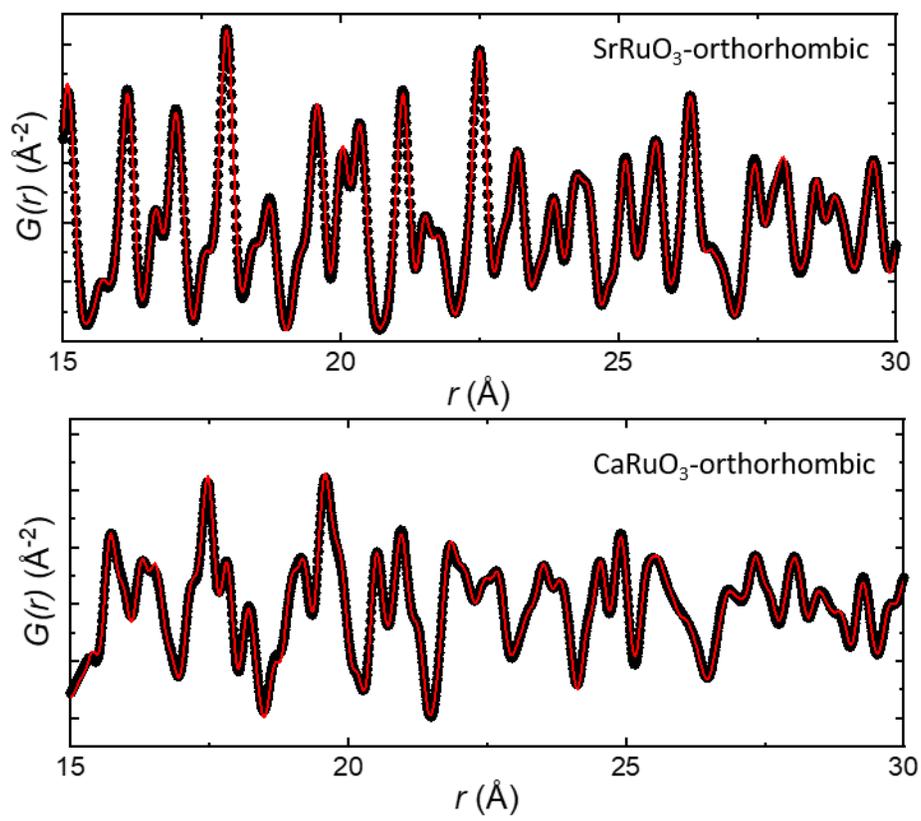

**Fig. 11**

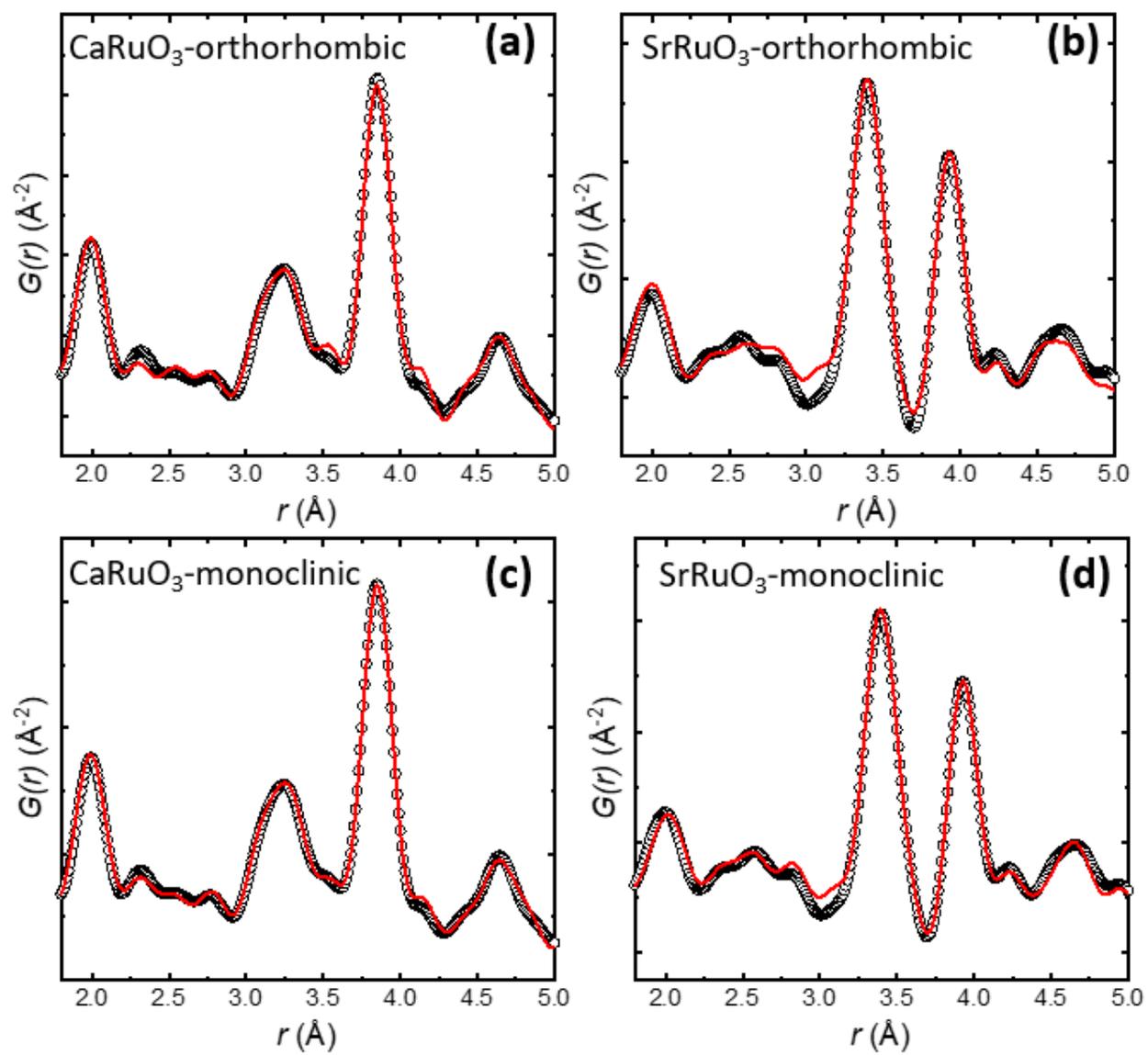

**Fig. 12**

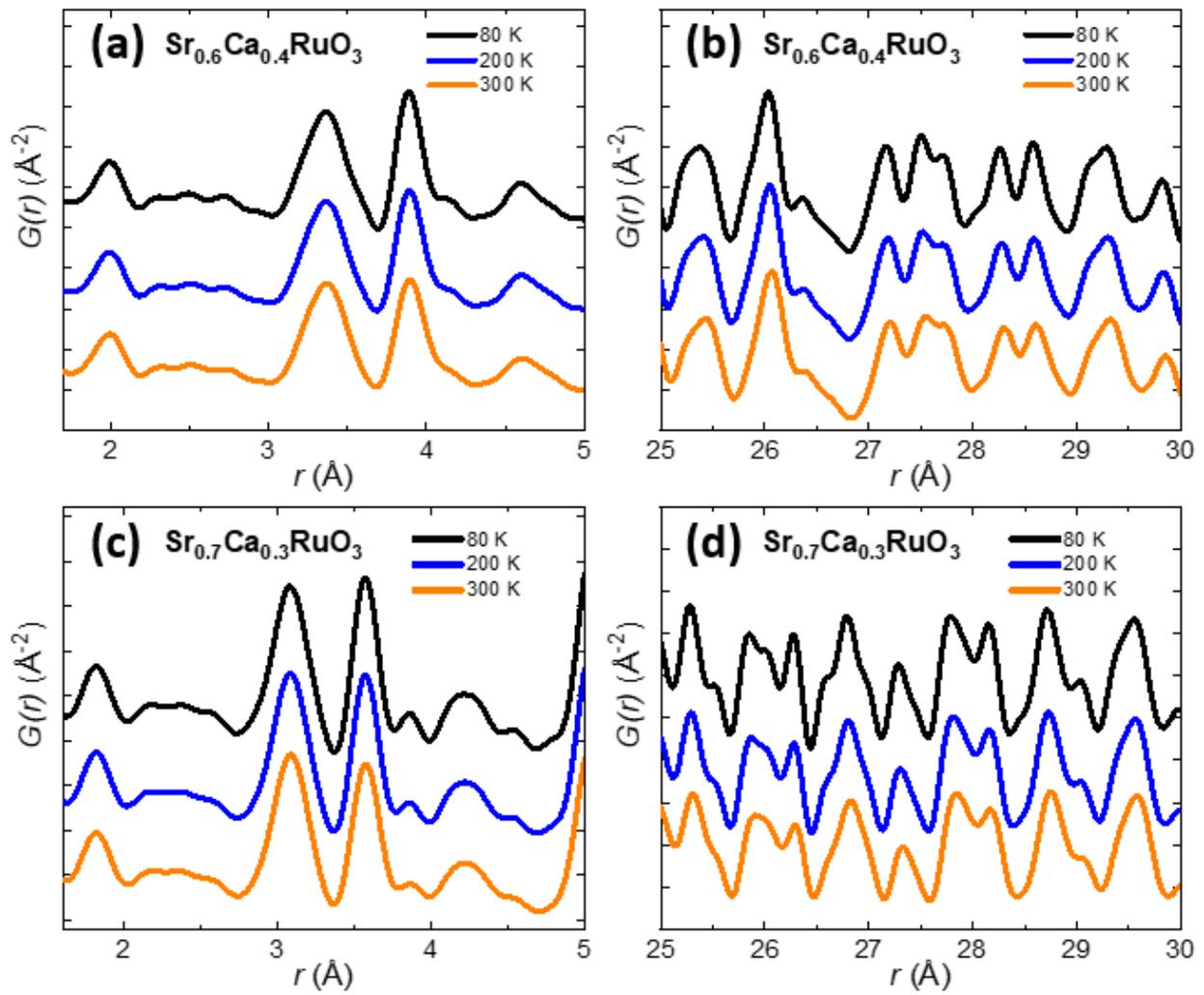

Fig. 13

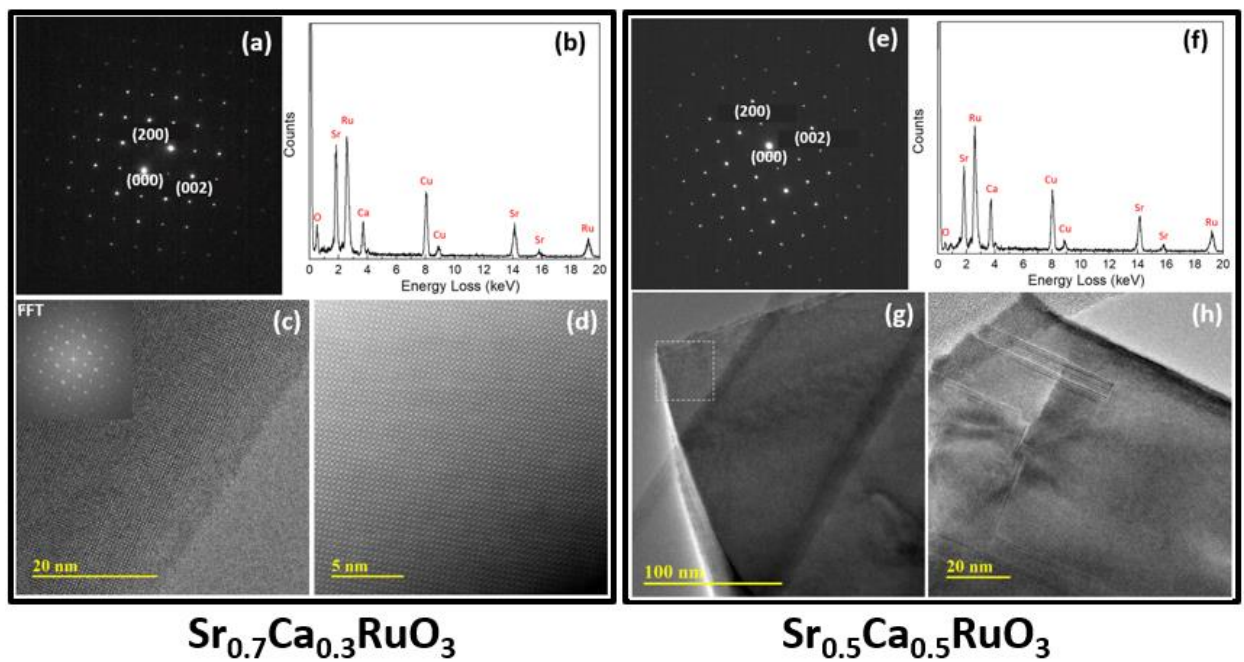

Fig. 14

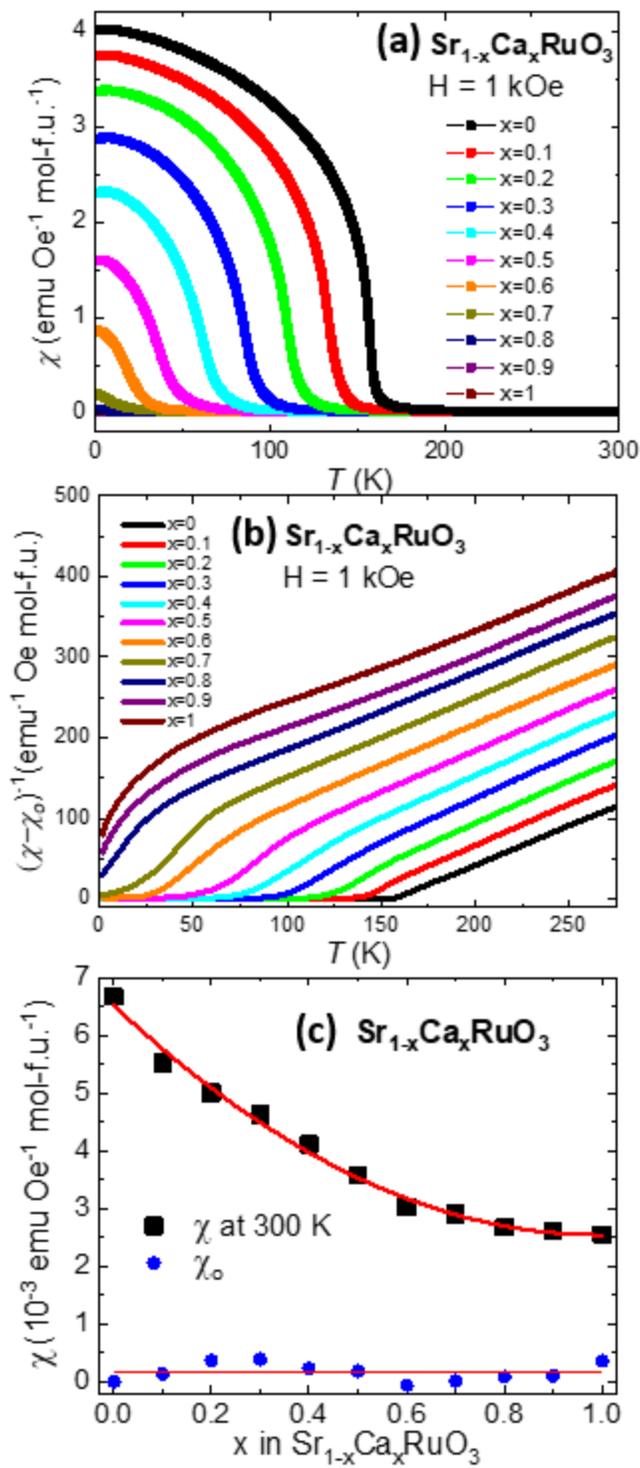

**Fig. 15**

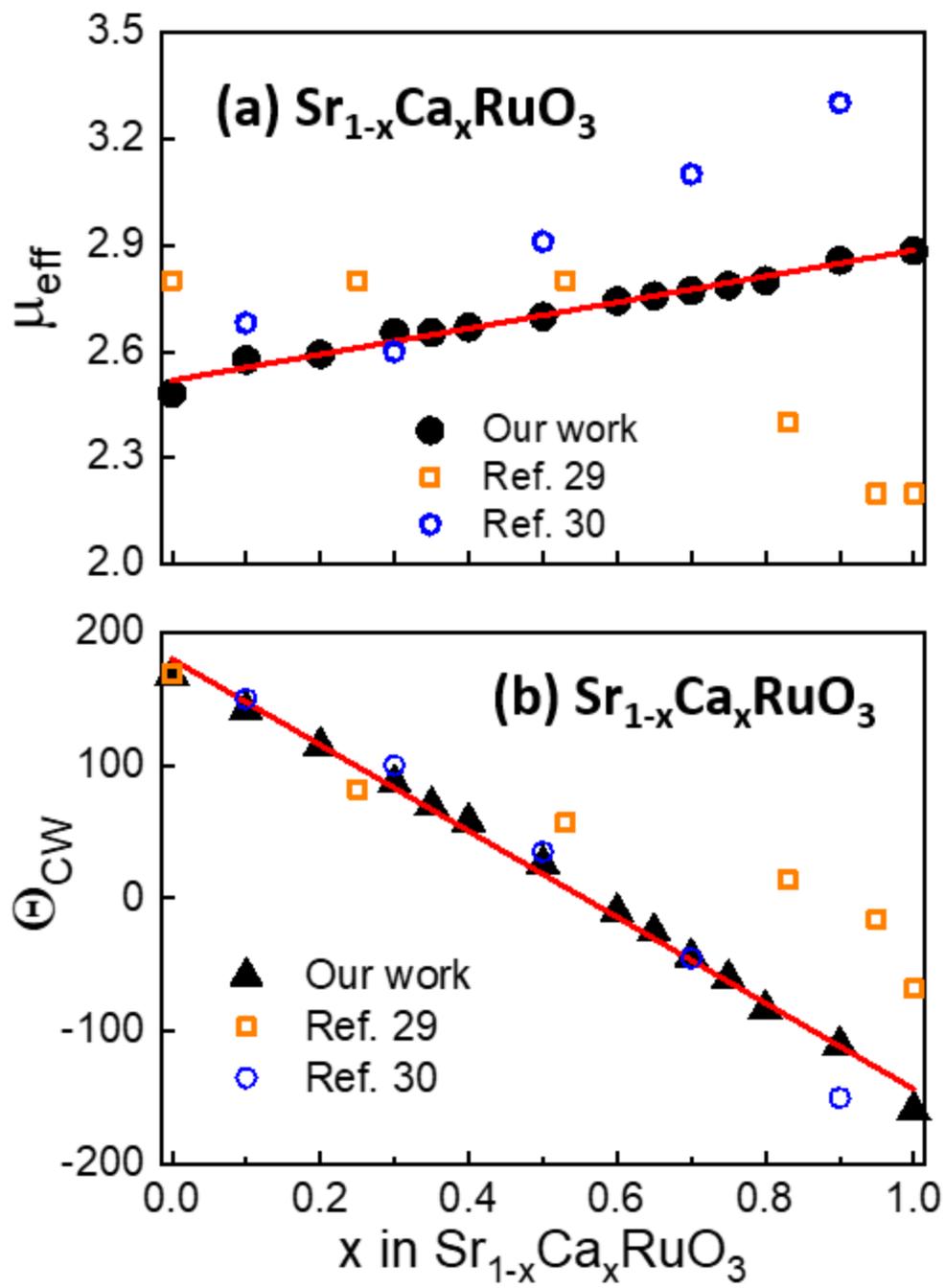

**Fig. 16**

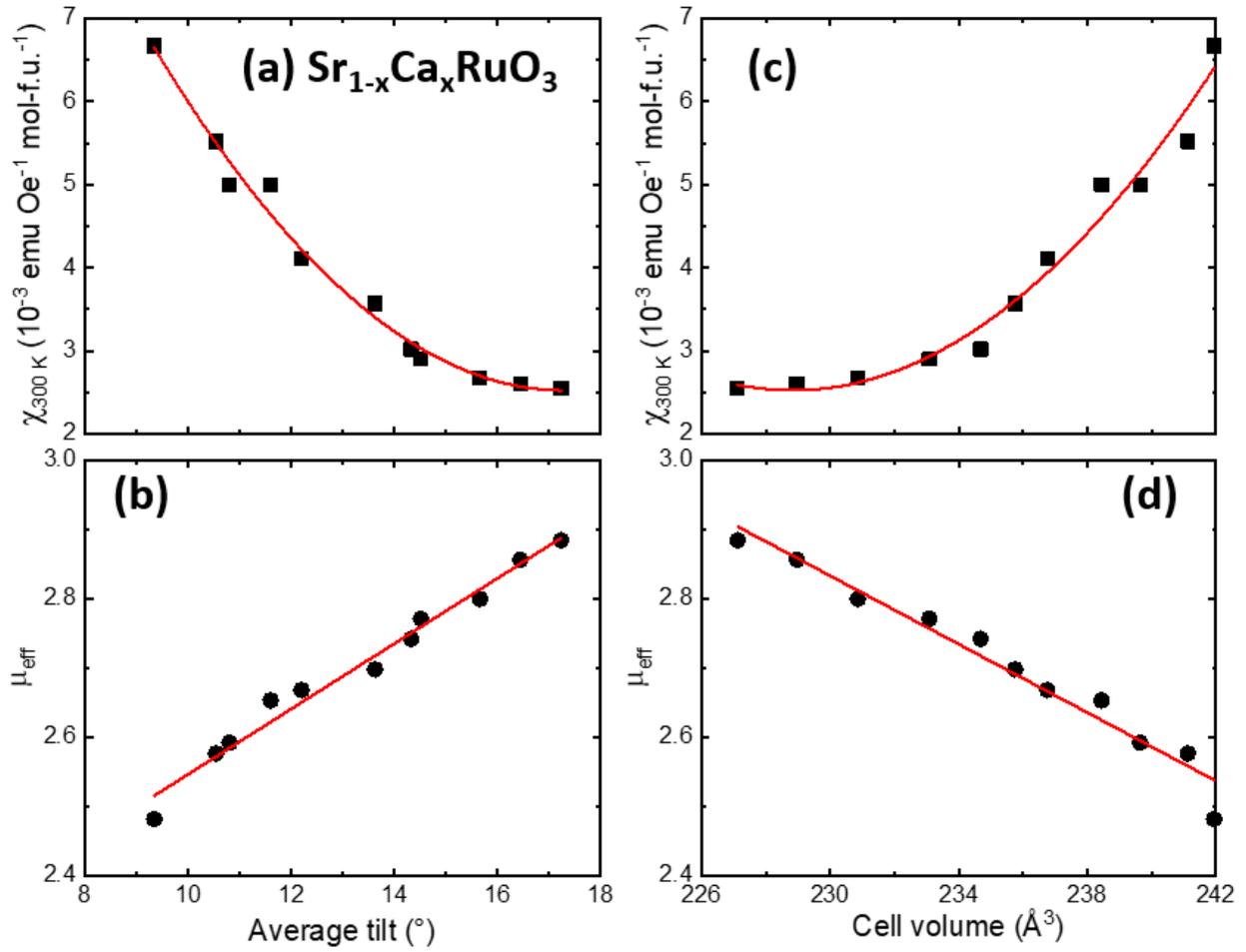

Fig. 17

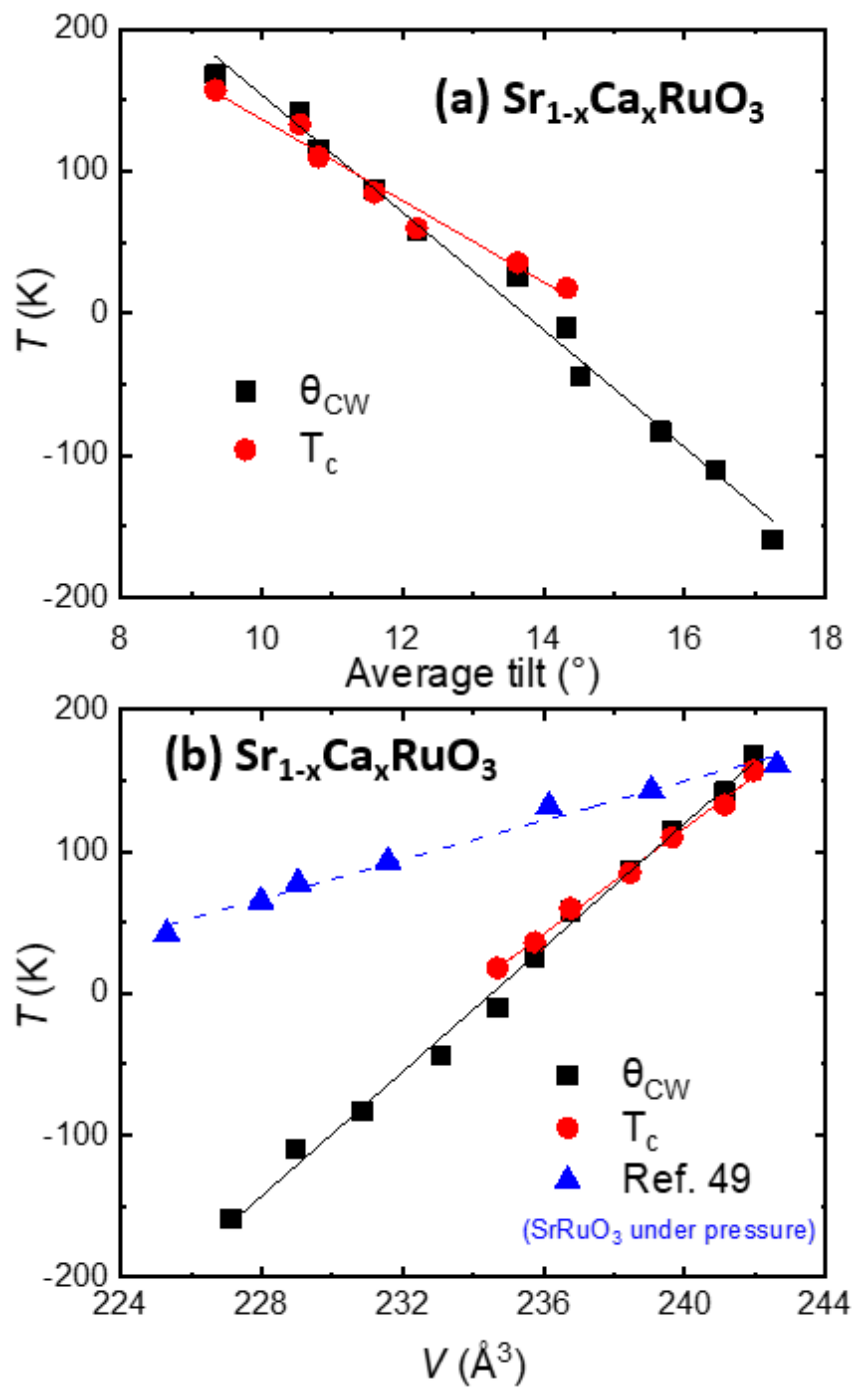

**Fig. 18**

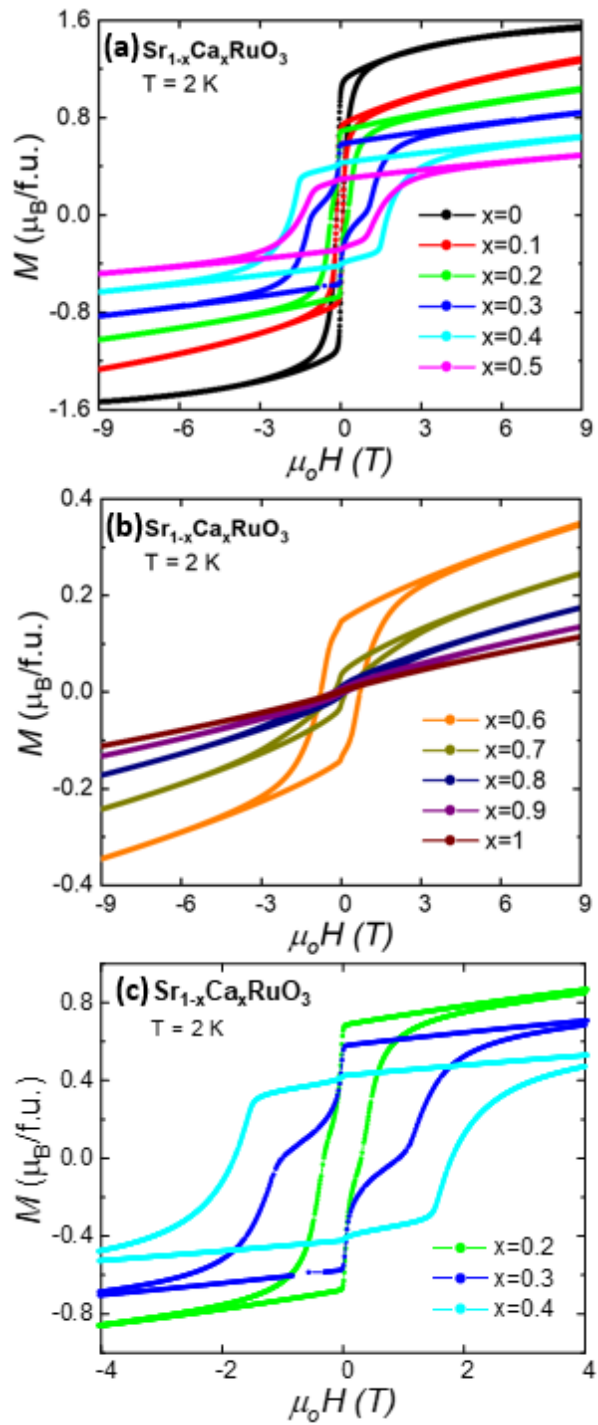

**Fig. 19**

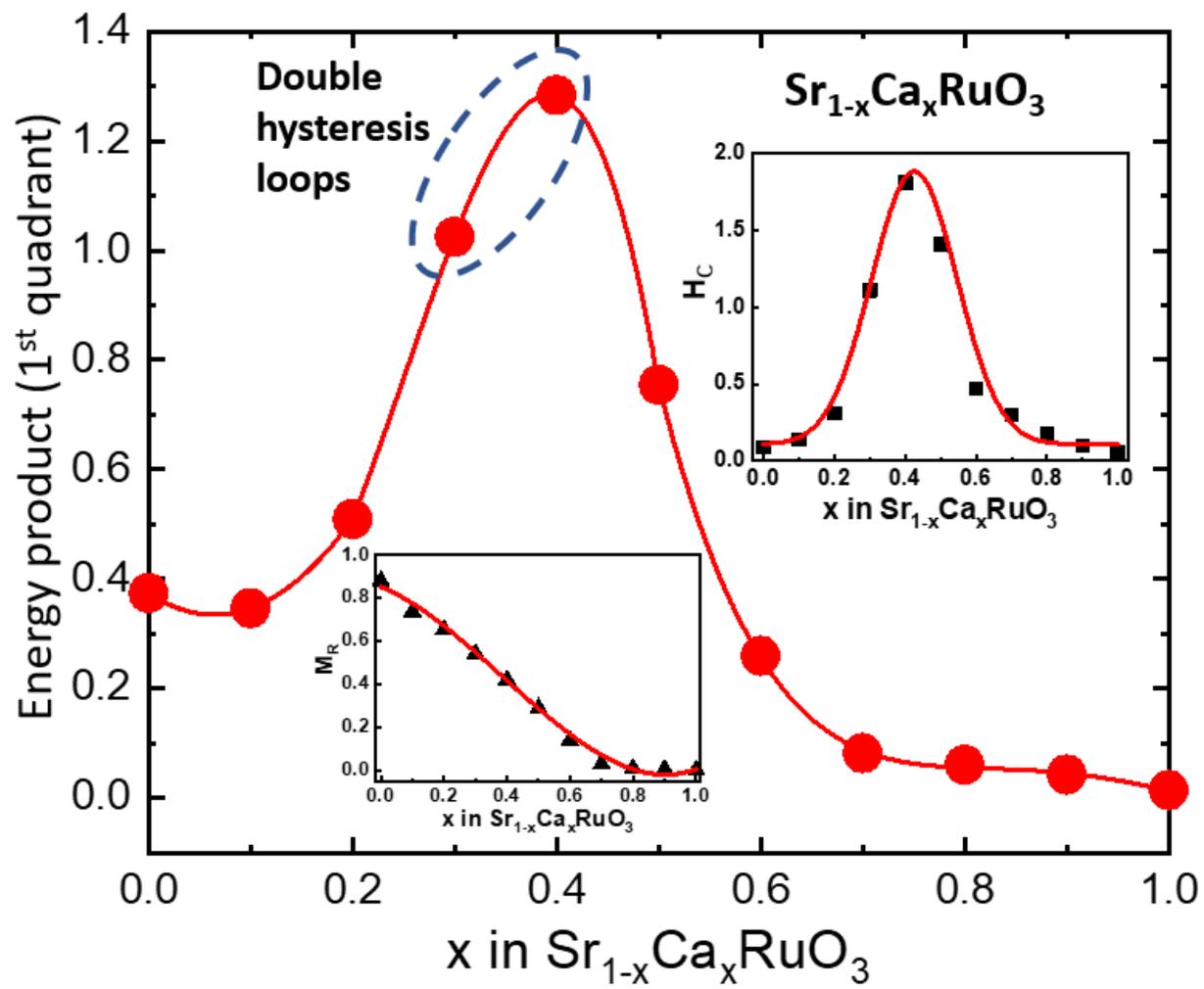

**Fig. 20**